\documentclass[conference,a4paper]{IEEEtran}

\usepackage{balance}
\usepackage[utf8]{inputenc} 
\usepackage[T1]{fontenc}
\usepackage{url}
\usepackage{ifthen}
\usepackage{cite}
\usepackage[cmex10]{amsmath} 
\usepackage{amssymb}
\usepackage{color}

\ifCLASSINFOpdf
  \usepackage[pdftex]{graphicx}
\else
  \usepackage[dvips]{graphicx}
\fi
\newcommand{\Fig}[1]{Fig.~\textup{\ref{#1}}}
\graphicspath{{figs/}}

\begin{document}
\title{On LDPC Code Based\\Massive Random-Access Scheme\\for the Gaussian Multiple Access Channel}

 \author{%
   \IEEEauthorblockN{Luiza Medova\IEEEauthorrefmark{1}\IEEEauthorrefmark{2},
                     Anton Glebov\IEEEauthorrefmark{3}, 
                     Pavel Rybin\IEEEauthorrefmark{1}, 
                     and Alexey Frolov\IEEEauthorrefmark{3}\IEEEauthorrefmark{1}}
   \IEEEauthorblockA{\IEEEauthorrefmark{1}%
                     Inst. for Information Transmission Problems,
                     Moscow, Russia,
                     prybin@iitp.ru}
   \IEEEauthorblockA{\IEEEauthorrefmark{2}%
                     Moscow Inst. of Physics and Mathematics,
                     Moscow, Russia,
                     luiza.medova@phystech.edu}
   \IEEEauthorblockA{\IEEEauthorrefmark{3}%
                     Skolkovo Institute of Science and Technology, 
                     Moscow, Russia,
                     \{anton.glebov@skolkovotech.ru, al.frolov@skoltech.ru\}}
 }

\maketitle

\begin{abstract}
This paper deals with the problem of massive random access for Gaussian multiple access channel (MAC). We continue to investigate the coding scheme for Gaussian MAC proposed by A. Vem et al in 2017. The proposed scheme consists of four parts: (i) the data transmission is partitioned into time slots; (ii) the data, transmitted in each slot, is split into two parts, the first one set an interleaver of the low-density parity-check (LDPC) type code and is encoded by spreading sequence or codewords that are designed to be decoded by compressed sensing type decoding; (iii) the another part of transmitted data is encoded by LDPC type code and decoded using a joint message passing decoding algorithm designed for the T-user binary input Gaussian MAC; (iv) users repeat their codeword in multiple slots. In this paper we are concentrated on the third part of considered scheme. We generalized the PEXIT charts to optimize the protograph of LDPC code for Gaussian MAC. The simulation results, obtained at the end of the paper, were analyzed and compared with obtained theoretical bounds and thresholds. Obtained simulation results shows that proposed LDPC code constructions have better performance under joint decoding algorithm over Gaussian MAC than LDPC codes considered by A. Vem et al in 2017, that leads to the better performance of overall transmission system.
\end{abstract}


\section{Introduction}

Current wireless networks are designed with the goal of servicing human users. Next generation of wireless networks is facing a new challenge in the form of machine-type communication: billions of new devices (dozens per person) with dramatically different traffic patterns are expected to go live in the next decade. The main challenges are associated with: (a) huge number of autonomous devices connected to one access point, (b) low energy consumption, (c) short data packets. This problem has attracted attention (3GPP and 5G-PPP) under the name of mMTC (massive machine-type communication).

There are $K \gg 1$ users, of which only $T$ have data to send in each time instant. A base station (BS) sends periodic beacons, announcing frame boundaries, so that the uplink (user-to-BS) communication proceeds in a frame-synchronized fashion. Length of each frame is $N$, where a typical interesting value is $n \approx 10^4 - 10^5$ . Each active user has $k$ bits that it intends to transmit during a frame, where a typical value is $k \approx 100$ bit. The main goal is to minimize the energy-per-bit spent by each of the users. We are interested in grant-free access (5G terminology). That is, active users transmit their data, without any prior communication with the BS (without resource requests). We will focus on the Gaussian multiple-access channel (GMAC) with equal-power users, i.e.
\[
Y = \sum\limits_{t=1}^{T}X_{t} + Z,
\]
where $Z \sim \mathcal{N}(0, N_0/2)$ and $\mathbb{E}\left[ |X_i|^2 \right] \leq P$.

This paper deals with construction of low-complexity random coding schemes for GMAC (indeed we restrict our consideration to the case of binary input GMAC). Let us emphasize the main difference from the classical setting. Classical information theory provided the exact solutions for the case of all-active users, i.e. $T = K$. Almost all well-known low-complexity coding solutions for the traditional MAC channel (e.g. \cite{rimoldi1996rate}) implicitly assume some form of coordination between the users. Due to the gigantic number users we assume them to be symmetric, i.e. the users use the same codes and equal powers. Here we continue the line of work started in \cite{polyanskiy2017perspective, OrdentlichPolyanskiy2017, Vem2017}. In \cite{polyanskiy2017perspective} the bounds on the performance of finite-length codes for GMAC are presented. In \cite{OrdentlichPolyanskiy2017} Ordentlich and Polyanskiy describe the first low-complexity coding paradigm for GMAC.  The improvement (it terms of required $E_b/N_0$) was given in \cite{Vem2017}. 

We continue to investigate the coding scheme from \cite{Vem2017}. The proposed scheme consists of four parts: 
\begin{itemize}
\item the data transmission is partitioned into time slots; 
\item the data, transmitted in each slot, is split into two parts, the first one (preamble) allows to detect users that were active in the slot. It also set an interleaver of the low-density parity-check (LDPC) type code \cite{Gallager63ldpc, tanner1981recursive} and is encoded by spreading sequence or codewords that are designed to be decoded by compressed sensing type decoding; 
\item the second part of transmitted data is encoded by LDPC type code and decoded using a joint message passing decoding algorithm designed for the $T$-user binary input GMAC; 
\item users repeat their codeword in multiple slots and use successive interference cancellation.  
\end{itemize}
The overall scheme can be called T-fold irregular repetition slotted ALOHA (IRSA, \cite{Liva, Pfister}) scheme for GMAC. The main difference of this scheme in comparison to IRSA is as follows: any collisions of order up to T can be resolved with some probability of error introduced by Gaussian noise. 

In this paper we are concentrated on the third part of considered scheme. Our contribution is as follows. We generalized the protograph extrinsic information transfer charts (EXIT) to optimize the protograph of LDPC code for GMAC. The simulation results, obtained at the end of the paper, were analyzed and compared with obtained theoretical bounds and thresholds. Obtained simulation results shows that proposed LDPC code constructions have better performance under joint decoding algorithm over Gaussian MAC than LDPC codes considered in \cite{Vem2017}, that leads to the better performance of overall system.

\section{Iterative joint decoding algorithm}

\begin{figure*}
\centering
\includegraphics[width=0.8\textwidth]{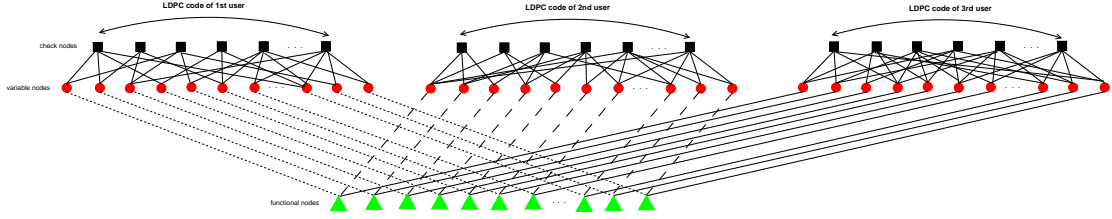}
\caption{Joint decoder graph representation for $T = 3$}
\label{fig:fg}
\end{figure*}

By $\mathcal{C}^{(t)}$, $t \in [T]$, we denote the codes used by users (the codes are binary). Recall, that $n$ and $R$ are accordingly the length and the rate of  $\mathcal{C}^{(t)}$, $k \in [T]$.

Thus $T$ users send codewords $\mathbf{c}^{(1)}, \mathbf{c}^{(2)}, \ldots, \mathbf{c}^{(T)}$. After BPSK modulator we have the sequences $\mathbf{x}^{(1)}, \mathbf{x}^{(2)}, \ldots, \mathbf{x}^{(K)}$, $\mathbf{x}^{(i)} \in \{-\sqrt{P}, +\sqrt{P}\}^n$. The channel output ($\mathbf{y}$) is the element-wise sum of the sequences affected by Gaussian noise.

The aim of joint multi-user decoder is to recover the all the codewords based on received vector $\mathbf{y}$. The decoder employs a low-complexity iterative belief propagation (BP) decoder that deals with a received soft information presented in LLR (log likelihood ratio) form. The decoding system can be represented as a graph (factor graph), which is shown in \Fig{fig:fg}. User LDPC codes are presented with use of Tanner graphs with variable and check nodes. At the same time there is a third kind of nodes in the figure -- functional nodes (marked with green color). These nodes correspond to the elements of the received sequence $\mathbf{y}$. 

Following the by now standard methodology of factor graphs, see \cite{richardson2008modern, kschischang2001factor}, we can write down the corresponding message passing decoding algorithm.

\begin{enumerate}
\item initialize the LLR values of variable nodes for each user code with zero values assuming equal probability for $1$ and $-1$ values;
\item perform $I_{O}$ outer iterations, where each iteration consists of the following steps:
\begin{enumerate}
\item perform maximum likelihood decoding of functional nodes (i.e. calculate update messages for variable nodes);
\item perform $I_{I}$ inner iterations of BP decoder for users' LDPC codes and update LLR values of variable nodes (this is done in parallel);
\end{enumerate}
\end{enumerate}

The BP part is standard, i.e. each user utilizes standard BP decoding algorithm (Sum-Product or Min-Sum) to decode an LDPC code. The most interesting part is the decoding of functional nodes. Following the principles of message-passing algorithms, the update rule to compute the message ($\mu$) sent to $i$-th variable node of $k$-th user ($k=1, \ldots, K,\:\: i=1, \ldots, N$) from a functional node $\mathcal{F}_i$ is the following:

Let $m_{vc,j} ^{k}$ denotes the message sent by the variable node $v$ to the check node $c$ along its $jth$ edge of user $k$:
\[
m_{vc,j} = \sum_{i = 1 ,~i \neq j}^{d_v - 1} m_{cv,i}^{k}  + m_{sv}^k ;
\]
where $m_{cv,i}^{k} $ is the message outgoing from the check node :
\[
m_{cv,j} = 2 \tanh^{-1} \left( \prod_{i = 1 ,~i \neq j}^{d_c - 1} \tanh \frac{m_{vc,i} }{2} \right) ;
\]
and $m_{sv}^k$ is the message outgoing from the state node.

It is also necessary to describe the rule for computing messages outgoing from state nodes.  Let $x_i^k$ denote the $i^{th}$ transmitted code bit and $y_i $ denote the channel output.  The outgoing message from the $i^{th}$ variable node of user $k$ to the connected state node is
computed as 
\[
m_{vs,i}^k = \log \frac{p(x_i^k = 1)}{p(x_i^k = -1)},~~ e^{m_{vs,i}^k} =  \frac{p(x_i^k = 1)}{p(x_i^k = -1)} ;
\]
Considering standard function node message-passing rules \cite{richardson2008modern}, we compute the message sent to user $k$ $ith$  variable node from the state node:
\[
m_{sv,i} ^{k} = \log  \frac{p(x_i^k = 1|y)}{p(x_i^k = -1|y)} =
\]
{
\small
\[
\log \left( \frac{ \sum\limits_{\sim x_i ^{(k)}} \prod\limits_{j \neq k} p(x_i^j = 1) p(y_i | x_i ^ {(1)}, ..., x_i ^ {(k)} = 1, ..., x_i ^ {(n)})}{ \sum\limits_{\sim x_i ^{(k)} } \prod\limits_{j \neq k}p(x_i^j = -1) p(y_i | x_i ^ {(1)}, ...,x_i ^{(k)} = -1, ...,  x_i ^ {(n)})}  \right)
\]
}
We can simplify it in the following way:
{
\small
\begin{flalign}\label{eq:m_sv}
 & m_{sv,i} ^{k} = \nonumber \\
& \log \left( \frac{\sum\limits_{\sim x_i ^{(k)}}  \prod\limits_{j \neq k}  e^{1_{x_j} X_j} p(y_i | x_i ^ {(1)}, ..., x_i ^ {(k)} = 1, ..., x_i ^ {(n)})}{\sum\limits_{\sim x_i ^{(k)}}   \prod\limits_{j \neq k} e^{1_{x_j} X_j}  p(y_i | x_i ^ {(1)}, ..., x_i ^ {(k)} = -1, ..., x_i ^ {(n)})}\right),
\end{flalign}
}
where $~~1_{x_k} = \begin{cases}

1, &\text{ $x_i ^{(j)} = ~1$}\\
0, &\text{ $x_i ^{(j)} = -1.$}
\end{cases}$.

The number of computations necessary to obtain the outgoing messages from the node $\mathcal{F}_{i}$ grows exponentially with the number of users, nevertheless, this number of users usually remains small, and we will therefore not be concerned with this fact.

\section{PEXIT Charts}

Extrinsic Information Transfer (EXIT) charts \cite{brink2001} can be used for the accurate analysis of the behavior of LDPC decoders. But since the usual EXIT analysis  cannot be applied to the study of protograph-based \cite{Thorpe2003} LDPC codes we will use a modified EXIT analysis for protograph-based LDPC codes (PEXIT) \cite{LivaChiani2007}. This method is similar to the standard EXIT analysis in that it tracks the mutual information between the message edge and the bit value corresponding to the variable node on which the edge is incident, while taking into account the structure of the protograph. In our work we use the notation from  \cite{LivaChiani2007} to describe EXIT charts  for protograph-based LDPC codes.
	
Let $ I_{Ev}$ denote the extrinsic mutual information between a message at the output of a variable node and the codeword bit associated to the variable node:
\[
I_{Ev} = I_{Ev}\left(I_{Av},I_{Es}\right),
\]
where $ I_{Av} $  is the mutual information between the codeword bits and the check-to-variable messages and $ I_{Es} $ is the mutual information between the codeword bits and the state-to-variable messages. Since the PEXIT tracks the mutual information on the edges of the protograph, we define $I_{Ev}(i,j)$ as the mutual information between the message sent by the variable node $V_j$ to the check node $C_i$ and the associated codeword bit:
\[ 
I_{Ev}(i,j) =  J\left( \sqrt{ \sum_{s \neq i}  [J^{-1}(I_{Av}(s,j))]^2 + [J^{-1}(I_{Es}(j))]^2 } \right) 
\]
where $J(\sigma)$ is given by \cite{brink2001}:
\begin{flalign*}
& J(\sigma)=1 & \\ &- \int\limits_{-\infty}^{\infty} \frac{1}{\sqrt{2\pi \sigma^2}} \exp\left[-\frac{1}{2}\left(\frac{y-\frac{\sigma^2}{2}x}{\sigma}\right)^2\right] \log_{2}(1+e^{-y})dy.&
\end{flalign*}

Similarly, we define $I_{Ec}$ , the extrinsic  mutual information between a message at the output of a check node and the codeword bit associated to the variable node receiving the message:
\[
I_{Ec}=I_{Ec} \left(I_{Ac}\right),
\]
where $I_{Ac}$ is the mutual information between one input message and the associated codeword bit and $I_{Ac}=I_{Ev}$. Accordingly, the mutual information between the message sent by $C_i$ to $V_j$ and the associated codeword bit is described as:
\[
I_{ec}(i,j) =1 -  J\left(\sqrt{\sum_{s \neq j} [J^{-1}(1-I_{ac}(i,s))]^2} \right).
\]

The mutual information between the variable node $V_j$ and the message passed to the state node is denoted as $I_{Evs}(j)$ and is given by:
\[
I_{Evs}(j) = J\left(\sqrt{\sum_{s} [J^{-1}(I_{av}(s,j))]^2} \right).
\]

Next we need to compute the mutual information $I_{Es}$. In order to get an idea about the probability density function of (\ref{eq:m_sv}) for user $j$, we generate samples of the outgoing LLRs through (\ref{eq:m_sv}) based on the samples of the received LLRs from other users whose PDF is approximated with  $ \mathcal{N}(\mu_{Evs} ,2 \mu_{Evs})$, where $\mu_{Evs} = \frac{J^{-1}(I_{Evs})}{2}$. To numerically estimate $\mu_{Es}$ and obtain the required mutual information as $I_{Es} = J(\mu_{Es})$, we refer to  \cite{Shahid2014}, where  the following three approaches are proposed:
\begin{itemize}
\item Mean-matched Gaussian approximation : the mean $\mu$ is estimated  from samples and we set $\mu_{Evs} = \mu$ and $\sigma^2_{Evs} = 2\mu$.
\item  Mode-matched Gaussian approximation : given  a sufficiently large number of $N$ samples generated through (\ref{eq:m_sv}), the mode $m$ is estimated  from samples  and we set $\mu_{Evs} = m$ and $\sigma^2_{Evs} = 2m$.
\item Gaussian mixture approximation: mean values $ \mu_1 , ...,\mu_k $ and the weights $a_1 ,..., a_k$ are estimated from samples and $I_{Es} = a_1J(\mu_1) + ... +  a_kJ(\mu_k).$
\end{itemize}

The rationale for using these approximations was shown in  \cite{Shahid2014}. Furthermore, the authors compared the performance of these approaches.  The mode-matched method was  found to give the maximum output mutual information and the joint codes designed by using this approximation also yield the lowest decoding bit error probability compared to the other two approaches.

Each user calculate $ I_{APP}(j)$, the mutual information between the  posteriori probability likelihood ratio evaluated by the variable node $ V_j $ and the associated codeword bit.
\[
I_{APP}(j) =  J\left( \sqrt{ \sum_{s }  [J^{-1}(I_{Av}(s,j))]^2 + [J^{-1}(I_{Es}(j))]^2 } \right).
\]
The convergence is declared if each $ I_{APP}(j)$ reaches 1 as the iteration number tends to infinity.

\section{Numerical Results}

\begin{figure}[htbp]
\centering
\includegraphics[width=\linewidth]{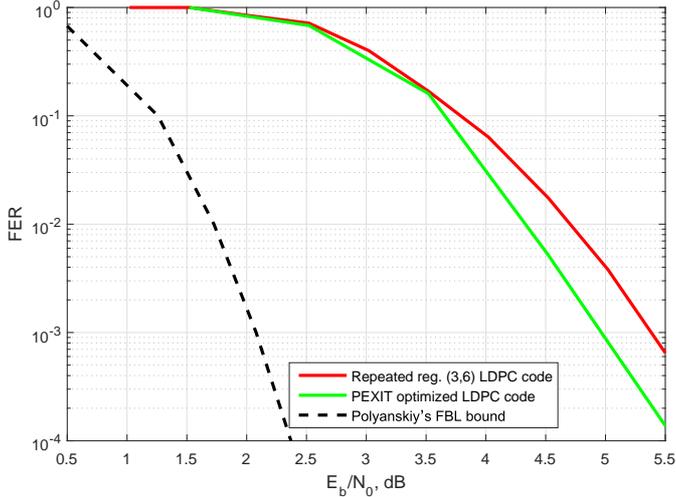}
\caption{Simulation results for T=2 and LDPC code (364, 91)}
\label{fig:compareT2}
\end{figure}

In this section the simulation results, obtained for the cases T=2 and T=4, are represented. Let us at first consider the simulation results for T=2 (Fig.~\ref{fig:compareT2}). For this case we compare the Frame Error Rate (FER) performance of rate-$1/4$ LDPC code (364, 91) from \cite{Vem2017} obtained by repetition of each code bit of regular (3,6) LDPC code twice, rate-$1/4$ LDPC code (364, 91) optimized by PEXIT charts method described above and Polyanskiy’s finite block length (FBL) bound for 2 user case.
As we can see in Fig.~\ref{fig:compareT2} proposed PEXIT-optimized LDPC code construction outperforms LDPC code construction from \cite{Vem2017} by about 0.5 dB. In the same time the gap between Polyanskiy’s FBL bound and PEXIT-optimized LDPC code is about 3 dB. But we would like to point out that used here Polyanskiy’s FBL bound is for Gaussian signal and not for Binary Phase-Shift Keying (BPSK) modulation, used for simulation. So, we believe that this gap will be reduced is FBL bound for BPSK modulation is used.

\begin{figure}[htbp]
\centering
\includegraphics[width=\linewidth]{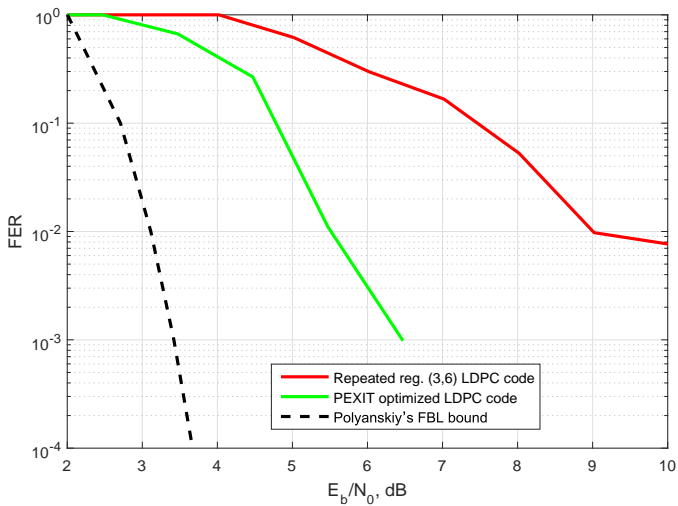}
\caption{Simulation results for T=4 and LDPC code (364, 91)}
\label{fig:compareT4}
\end{figure}

Now let us consider simulation results for T=4 (Fig.~\ref{fig:compareT4}). For this case we obtain another PEXIT-optimized rate-$1/4$ LDPC code (364, 91) and compare FER performance of same LDPC code from \cite{Vem2017} and Polyanskiy’s FBL bound for 4 users.
As we can see in Fig.~\ref{fig:compareT4} proposed PEXIT-optimized LDPC code construction outperforms LDPC code construction from \cite{Vem2017} by more than 3 dB. And again the gap between Polyanskiy’s FBL bound and PEXIT-optimized LDPC code is a little bit less than 3 dB.

\section{Sparse spreading of LDPC codes}

In this section we answer a very natural question: how to increase the order of collision, that can be decoded in a slot. E.g. consider the case from the previous section. Let the slot length $n' = 364$. We want to increase $T$ up to $8$. Here we face with two problems:
\begin{itemize}
\item The performance of LDPC joint decoder rapidly becomes bad with grows of $T$. We were not able to find $(364, 91)$ LDPC codes, that work well for $T = 8$. 
\item The number of computations necessary to obtain the outgoing messages from the functional node grows exponentially with the number of users $T$. 
\end{itemize}

\begin{figure}[htbp]
\centering
\includegraphics[width=\linewidth]{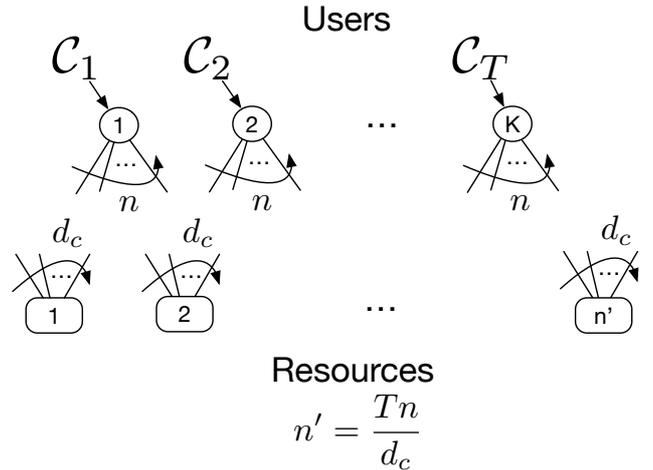}
\caption{Sparse spreading of LDPC codes}
\label{fig:spreading}
\end{figure}

We address both these problems in a scheme, which is proposed below (see \Fig{fig:spreading}). The idea is to use sparse spreading signatures \cite{LDS} for LDPC codes, such that the degree of functional node is reduced from $T$ to $d_c$. The slot length is now $n'$, $n' \ne n$. 

\begin{figure}[htbp]
\centering
\includegraphics[width=\linewidth]{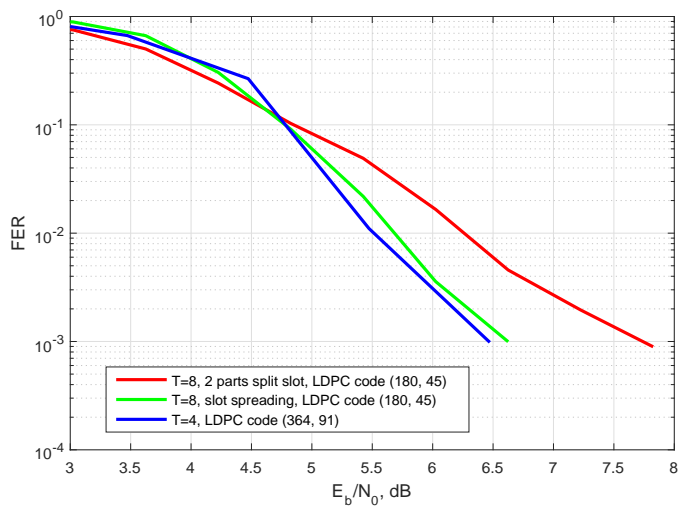}
\caption{Simulation results for spreading}
\label{fig:compareSpreading}
\end{figure}

In \Fig{fig:compareSpreading} we present the simulation results. As we were not able to find $(364, 91)$ LDPC codes, that work well for $T = 8$ we consider $2$ times shorter LDPC codes and compare $2$ strategies:
\begin{itemize}
\item split the slot into $2$ parts and send $4$ users in each part;
\item use sparse spreading;
\end{itemize} 

We see, that our approach is much better and works practically the same in comparison to the case of $2$ times longer LDPC codes and $2$ times smaller number of users (see the previous section).

\section{Conclusion}
We generalized the protograph extrinsic information transfer charts (EXIT) to optimize the protograph of LDPC code for GMAC. The simulation results, obtained at the end of the paper, were analyzed and compared with obtained theoretical bounds and thresholds. Obtained simulation results shows that proposed LDPC code constructions have better performance under joint decoding algorithm over Gaussian MAC than LDPC codes considered  by A. Vem et al in 2017, that leads to the better performance of overall system.

\section*{Acknowledgment}

We want to thank Y.~Polyanskiy for fruitful discussions.



\end{document}